\documentclass{INTERSPEECH2023}
\usepackage{amsthm,amsmath,amssymb}
\usepackage{multirow}
\usepackage{amsmath}
\usepackage{graphicx}
\usepackage{makecell}

\interspeechcameraready

\title{Ordered and Binary Speaker Embedding}
\name{Jiaying Wang$^{1,3}$, Xianglong Wang$^3$, Namin Wang$^2$, Lantian Li$^{3*}$, Dong Wang$^{1*}$\thanks{
This work was supported by the National Natural Science Foundation of China under Grant No.62171250,
and also the Huawei Cloud Research Program under the project No.TC20220615035.}}
\address{
  $^1$Center for Speech and Language Technologies, BNRist, Tsinghua University, China  \\
  $^2$Huawei Cloud, China~~~~~$^3$Beijing University of Posts and Telecommunications, China}
\email{$^*$Corresponding authors:~lilt@bupt.edu.cn, wangdong99@mails.tsinghua.edu.cn}

\begin{document}

\maketitle

\begin{abstract}

Modern speaker recognition systems represent utterances by embedding vectors.
Conventional embedding vectors are dense and non-structural.
In this paper, we propose an ordered binary embedding approach that sorts the dimensions
of the embedding vector via a nested dropout and converts the sorted vectors to binary codes via Bernoulli sampling.
The resultant ordered binary codes offer some important merits such as hierarchical clustering, reduced memory usage, and fast retrieval.
These merits were empirically verified by comprehensive experiments on a speaker identification task with the VoxCeleb and CN-Celeb datasets.

\end{abstract}

\section{Introduction}
\label{sec:intro}

Speaker recognition is the process of recognizing the identity of a person from his/her voice.
Due to its traits of being user-friendly, non-intrusive, non-touching, and low privacy,
speaker recognition has found broad real-life applications~\cite{zheng2017robustness}.
Modern speaker recognition systems are mostly based on the concept of `embedding',
i.e., represent a variable-length utterance by a fixed-dim dense vector.
Traditional speaker embedding is based on statistical models, in particular the i-vector model~\cite{dehak2010front}.
Recently, deep neural nets become the most popular embedding models~\cite{ehsan14,li2017deep,snyder2018x,cai2018exploring}
and achieved state-of-the-art performance in various benchmarks~\cite{sadjadi20202019,sadjadi20222021,nagrani2020voxsrc,brown2022voxsrc}.
Among all the embedding models, the x-vector model achieved the most success~\cite{snyder2018x}.

Despite the broad success of deep speaker embedding, the dense embedding vectors are not suitable for
large-scale identification tasks. For instance, per our experiment, identifying a person from 1,251 candidates
using 32-dim x-vectors costs 50 ms on a 1.2 GHz CPU and with the highly optimized Scipy package.
This amounts to 15.5 hours of CPU time per query if the size of candidates is 1.4 billion, the population of China.
This is unacceptable for most real applications.
To respond to this challenge, the recent CNSRC evaluation has set a large-scale identification task and required the participants to
report the time cost for search~\cite{wang2022cnsrc}.

A known approach to accelerating the search speed is by hierarchical clustering~\cite{apsingekar2009speaker,hu2012fuzzy}.
By this approach, enrolled speakers are clustered according to their similarity and multiple-level clustering forms a
decision tree. This approach largely reduces the search time, though the depth of the tree must be carefully controlled to avoid
a substantial performance drop.
Another direction is to use binary codes~\cite{jeon2012efficient,li2016binary}.
This approach converts dense embeddings to binary codes, and the similarity is computed as the Hamming distance which is much faster than
computing the cosine distance. The shortage of this approach is that the binarization process may lose precision, and the search still needs to scan all the candidates.

\begin{figure}[!htp]
  \centering
  \includegraphics[width=0.42\textwidth]{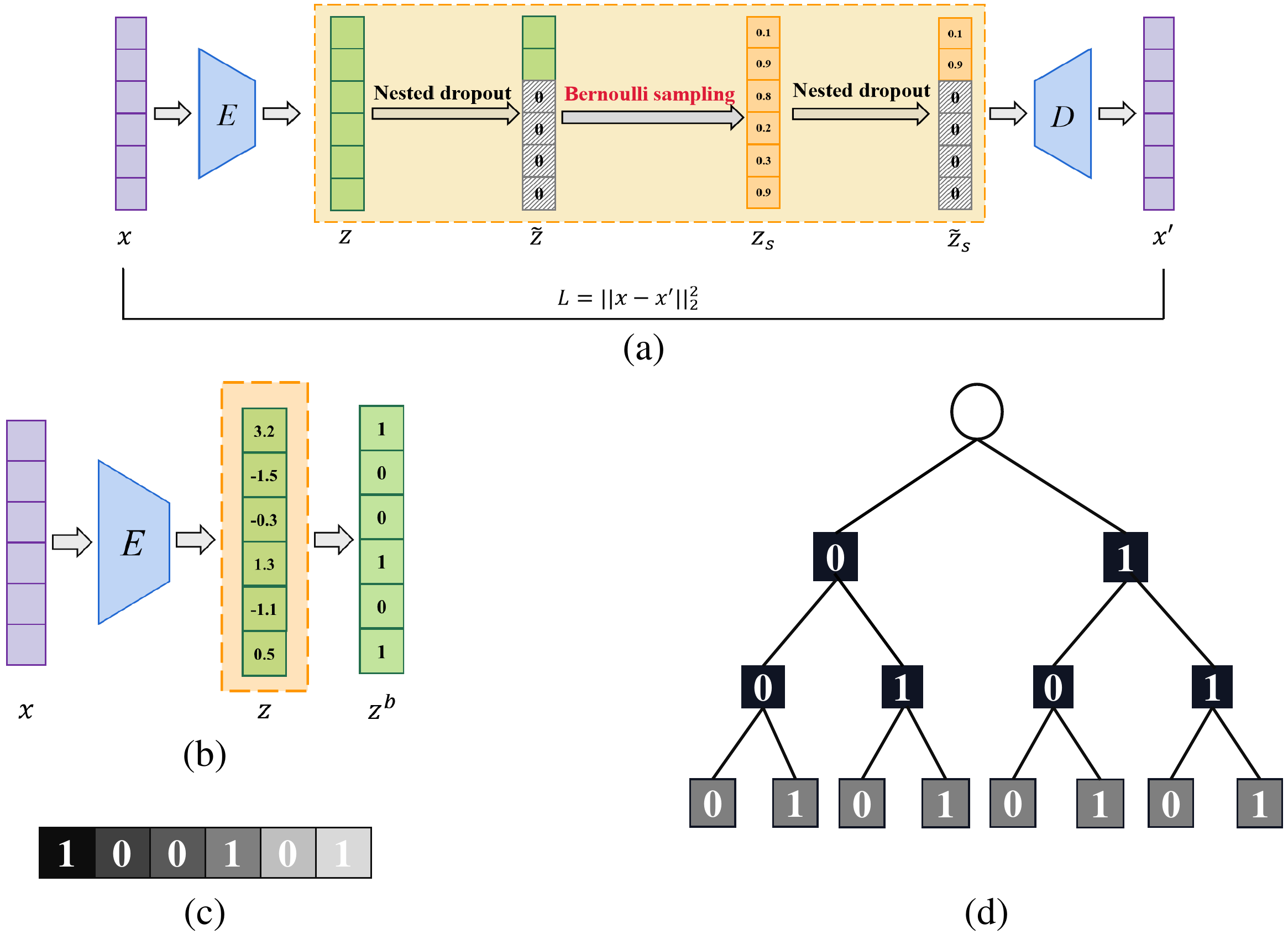}
  \vspace{-1mm}
  \caption{(a) The obAE model in the training phase. Details are in Section~\ref{sec:binary}.
  (b) The obAE model in the inference phase. The dense vector $x$ is converted to logits $z$ by the encoder, which is then converted to an OB code $z^b$.
  (c) An OB code example, with the grey level indicating the importance; (d) Binary tree formed by OB codes. The top circle is the starting point of the search.}
  \label{fig:main}
\end{figure}

In this paper, we propose a novel approach that involves the advantage of both hierarchical clustering and binary codes.
We design an ordered binary auto-encoder (obAE) model whose encoder
converts a dense embedding to an ordered binary (OB) code, where the bits of the code are ordered according to their importance for the
decoder to recover the input dense embedding. The architecture is shown in Figure~\ref{fig:main}(a), and the main architecture
involves a nested dropout~\cite{rippel2014learning} and a Bernoulli sampling.
Once the model is trained, the encoder can be used to produce OB codes, as shown in Figure~\ref{fig:main}(b)(c).
The implementation is as simple as several lines of Python code, and the additional computation cost is negligible.
More details will be presented in Section~\ref{sec:binary}.

\vspace{-0.5mm}
\vspace{-1mm}

On the one hand, OB codes are efficient in both search speed and memory usage, the same merit shared by other binary codes 
such as those derived by local sensitive Hashing (LSH)~\cite{gionis1999similarity}. However, since the OB code here is `learned' to minimize the reconstruction error,
it is more task-oriented, under the assumption that the reconstruction error is consistent with the similarity used in the search task, which is the case for speaker recognition as we will show in the experiments.
On the other hand, since the bits are ordered in OB codes, it naturally forms a binary tree that hierarchically clusters the enrolled speakers, as shown in Figure~\ref{fig:main}(d).
This offers a fast retrieval by a top-down tree traversal, a merit shared with the conventional hierarchical clustering approach~\cite{hu2012fuzzy}.
An advantage of the binary tree, however, is that the tree structure is optimized jointly with the code, leading to a better
accuracy-efficiency trade-off. Moreover, searching over the binary tree is just a sequence of bit matching, which is much faster than the
conventional tree search where at each node the query embedding needs to be compared with all the cluster centroids.


To verify the advantage of the OB, we conducted comprehensive experiments
on speaker identification (SID) tasks with the VoxCeleb~\cite{nagrani2020voxceleb} and CN-Celeb~\cite{li2022cn} datasets.
To our best knowledge, this paper is the first attempt to investigate ordered and binary speaker embedding.

\section{Related work}

Making features ordered is a long-standing topic. 
Principal components analysis (PCA) is perhaps the most popular approach~\cite{abdi2010principal}.
Since the emergence of deep models, multiple research has been conducted to impose dimensional orderliness in the latent space of
neural nets, especially auto-encoders (AEs).
Rippel et al.~\cite{rippel2014learning} proposed a nested dropout algorithm.
That applies structured masks on the latent features to encourage leading dimensions to take more important role.
Ladjal et al.~\cite{ladjal2019pca} proposed a PCA-like AE. They designed
an iterative training strategy that progressively increases the latent dimension, so that leading dimensions are 
learned first and more sufficiently. 
The merits of ordered features have been in multiple fields such as image compression, retrieval, and generation~\cite{xu2021anytime,shen2018ordered,lu2021progressive,cui2020fully}.
As far as we know, there is no investigation on ordered speaker embedding.

For binary speaker code,
\cite{jeon2012efficient} presented a binarization approach based on kernel LSH within the GMM-UBM framework.
Li et al.~\cite{li2016binary} presented an approach based on LSH as well, but performed the test on i-vectors.
Both studies reported reduced storage/memory and improved search speed with little or no loss of accuracy.

For ordered binary codes,
Rippel et al.~\cite{rippel2014learning} reported a simple `thresholding' approach, that cats a dense vector to binary codes according to a
predefined threshold. Xu et al.~\cite{xu2021anytime} presented a discretization approach based on K-means clustering
as in VQ-VAE~\cite{van2017neural}. Our obAE model differs from the above studies in that we obtain binary codes with Bernoulli sampling,
and train the model with the reparametrization trick~\cite{kingma2014stochastic}. 
We found this approach is simple in implementation and the training is stable.

\section{Method}

\subsection{Ordered AE (oAE)}

By setting a reconstruction loss and an appropriate information bottleneck~\cite{tishby2000information,tishby2015deep},
AE can discover a low-dimensional latent space that retains the most important information for describing the data.
Compared to PCA, AE is a non-linear model and is more flexible. The cost of the flexibility,
however, is that the latent space loses the property of dimensional orderliness.
Nested dropout on the latent features~\cite{rippel2014learning,xu2021anytime} can recover the order, leading to an ordered AE (oAE).
Note that oAE has been proposed in previous studies~\cite{rippel2014learning,xu2021anytime}, though not investigated in speaker recognition.


\begin{figure}[!htp]
  \centering
  \includegraphics[width=\linewidth]{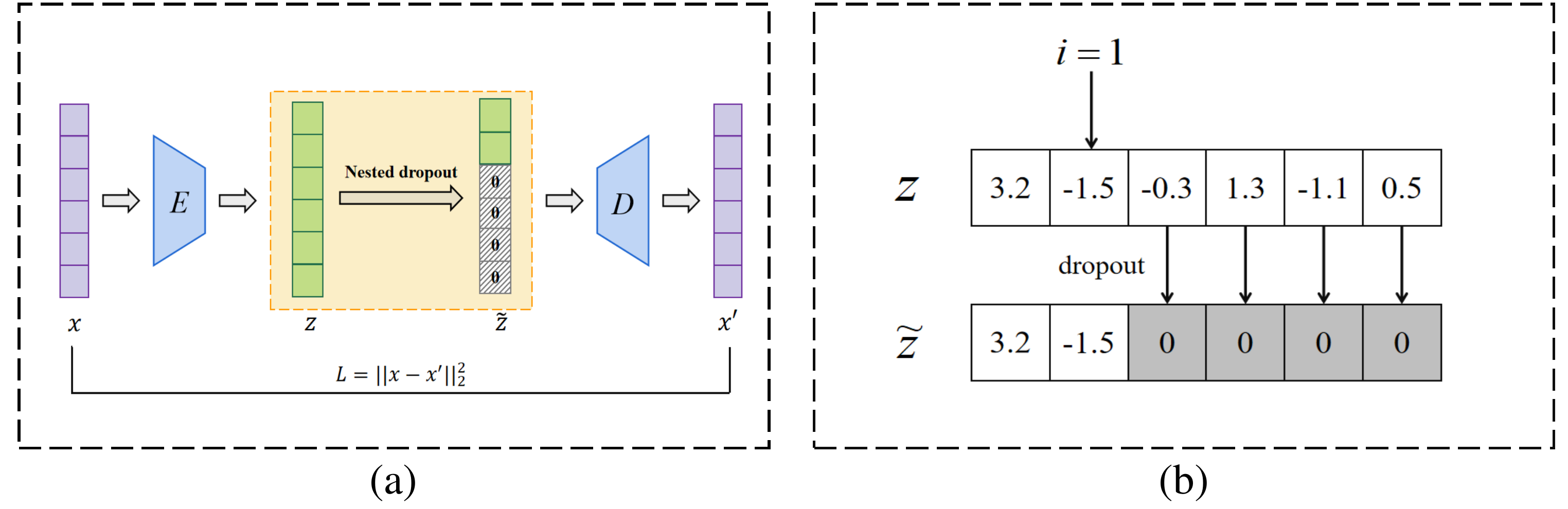}
  \vspace{-5mm}
  \caption{Illustration of (a) oAE architecture and (b) nested dropout, where $z$ and $\tilde{z}$ represent the latent features before and after dropout.
  Note that \textbf{all} the dimensions after the $i$-th dim are dropped out, i.e., set to 0. This is the main difference between nested dropout and conventional unit dropout.}
  \label{fig:nest}
\end{figure}

The oAE architecture and nested dropout are illustrated in Figure~\ref{fig:nest}.
Let $z$ denote the latent feature of an input data $x$. Nested dropout samples a dimension index $i$ following a uniform distribution,
and then masks all the dimensions of $z$ after the $i$-th dim, i.e., set them to zeros. The masked feature, denoted by $\tilde{z}=[z_{\le i}, 0,..,0]$, 
is then used to produce $x'$, the reconstruction of the input $x$. The model is trained with the MSE loss, as usual AEs.

Theoretical analysis~\cite{rippel2014learning,xu2021anytime} shows that the \emph{incremental} value of
the mutual information between $x$ and $z_{\le i}$, i.e., $\text{MI}(x, z_{\le i}) - \text{MI}(x, z_{\le i-1})$, declines
when $i$ increases. This means that $z_i$ is ordered according to the information it contains regarding the input, hence the 
importance for its reconstruction.
Another theoretical result is that oAE recovers PCA when the encoder and decoder are linear and share parameters~\cite{rippel2014learning}.
This establishes a more close link between PCA and AE models.

\subsection{Ordered binary AE (obAE)}
\label{sec:binary}

\subsubsection{Architecture}

We extend ordered AE to ordered binary AE by introducing a Bernoulli sampling. The model architecture has been shown in Figure~\ref{fig:main}(a).
Briefly, we treat the masked feature $\tilde{z}$ as the parameter of a Bernoulli distribution and sample a binary
code $z_s$ from it. This is formally written by $z_{s} \sim \text{Bernoulli}(p)$ where $p=\sigma(\tilde{z})$, with $\sigma(\cdot)$ the sigmoid function.
The sampled $z_s$ is then masked following the same way as $\tilde{z}$, producing
a masked binary code $\tilde{z}_s$. Finally, $\tilde{z}_s$ passes the decoder and the reconstructed input $x'$ is produced.
Again, the training is based on MSE loss.

In the test phase, both the masking and the Bernoulli sampling are not required, and we simply use the following
hash functions to obtain the OB code $z^b$:
\begin{equation}
z_i^b = \left\{
\begin{array}{rcl}
1     &      \ z_i \geq 0 \\
0     &      \ z_i \textless 0 \\
\end{array} \right.
\label{eq1}
\end{equation}

\subsubsection{Relaxed Bernoulli sampling}

A difficulty when training the obAE model is that the gradient cannot be back-propagated to the
encoder as $z$ impacts the output via a Bernoulli sampler. We solve the problem by 
using the resampling trick as in VAE~\cite{kingma2014stochastic}.
Specifically, we first sample $u$ from a uniform distribution on [0,1], and then produce $z_s$ as follows:
\begin{equation}
z_{si} = \left\{
\begin{array}{rcl}
1     &      \ u_i \leq p_i \\
0     &      \ u_i > p_i \\
\end{array} \right.
\label{eq:bern}
\end{equation}
\noindent where $p=\sigma(\tilde{z})$.

It is easy to verify that $z_s$ generated in this way follows $\text{Bernoulli}(p)$, and $\tilde{z}$ can receive gradients as it is not related to the sampler.
However, Eq. (\ref{eq:bern}) is a step function so cannot pass gradient. A sigmoid function can be used as a relaxed version, leading to the following procedure:
\begin{equation}
  z_s = \sigma \left(\frac{\log(\frac{u}{1-u}) + \log (\frac{p}{1-p})}{T} \right),
  \label{eq:sample}
\end{equation}
where $T$ controls the sharpness of the distribution of $z_s$.
The smaller the value $T$, the higher the probability that $z_s$ concentrates around 0 or 1.
In the limit case $T\rightarrow 0$, the distribution of $z_s$ approaches $\text{Bernoulli}(p)$.
In our experiments, we set $T$ to 0.1, and implement this sampling by the \emph{RelaxedBernoulli}($\cdot$) function supported by PyTorch\footnote{https://pytorch.org/docs/stable/distributions.html\#relaxedbernoulli}.


\section{Experiments}

In this section, we test the OB codes on speaker identification tasks with VoxCeleb~\cite{nagrani2020voxceleb} and CN-Celeb~\cite{li2022cn} datasets.
We first present the data and setups, and then report the results with ordered dense features and ordered binary codes in sequence.

\subsection{Data}

Two datasets were used in our experiments: VoxCeleb~\cite{nagrani2020voxceleb} and CN-Celeb~\cite{li2022cn}, and their information
is presented in Table~\ref{tab:data}.
It is worth noting that for VoxCeleb1, we selected the longest three utterances of each speaker for enrollment and the remaining for test.
For CN-Celeb.E, to avoid possible annotation errors in short utterances, we removed the test utterances whose duration is less than 2 seconds.

\begin{table}[htp!]
  \centering
  \caption{Data description}
  \vspace{-2.5mm}
  \label{tab:data}
  \scalebox{0.73}{
  \begin{tabular}{l|c|cc}
    \toprule
     \textbf{VoxCeleb}  & Train           &  Enroll       &  Test           \\
                        & VoxCeleb2.dev   &  VoxCeleb1    &  VoxCeleb1      \\
    \midrule
    \# of Spks          &  5,994          &  1,251        &  1,251          \\
    \# of Utters        &  1,092,009      &  3,753        &  149,763        \\
    \midrule
    \midrule
     \textbf{CN-Celeb}  & Train           &  Enroll       &  Test           \\
                        & CN-Celeb.T      &  CN-Celeb.E   &  CN-Celeb.E     \\
    \midrule
    \# of Spks          &  2,793          &  196          &  196            \\
    \# of Utters        &  632,740        &  196          &  14,124         \\
  \bottomrule
  \end{tabular}}
  \vspace{-4mm}
\end{table}

\subsection{Settings}

We use two publicly available x-vector models released in the Sunine toolkit\footnote{https://gitlab.com/csltstu/sunine}:
one was trained on VoxCeleb2.dev, and the other was trained on CN-Celeb.T.
The structure of both models is ResNet34~\cite{he2016deep} with squeeze-and-excitation (SE) layers~\cite{hu2018squeeze},
and the dimensionality of the x-vectors is 256.
We use the VoxCeleb model and the CN-Celeb model to extract x-vectors of the utterances in VoxCeleb and CN-Celeb, respectively.

For oAE and obAE models, the encoder and the decoder are both linear, though their parameters are not shared. We choose this simple structure for a direct comparison with PCA.
The source code is available online\footnote{https://github.com/AlexGranger-scn/OAE} to help readers reproduce our experiments.

\subsection{Baseline}

Firstly, the Top-k accuracy with the original x-vector (256 dims, or equally 8,192 bits) is reported in Table~\ref{tab:base}.
It can be observed that although the number of speakers in CN-Celeb.E is much smaller than that in VoxCeleb1,
the Top-k accuracies on CN-Celeb.E are clearly inferior to those on VoxCeleb1, indicating that CN-Celeb.E is a more challenging dataset for SID.

\begin{table}[!htp]
  \caption{Top-k accuracy with the original x-vector.}
  \vspace{-2mm}
  \centering
  \label{tab:base}
  \scalebox{0.75}{
  \begin{tabular}{c| c| c}
    \toprule
    \textbf{} & \textbf{VoxCeleb1}  &  \textbf{CN-Celeb.E}   \\
    \midrule
           Top-1        &  0.959              & 0.706   \\
           Top-3        &  0.984              & 0.800   \\
           Top-5        &  0.989              & 0.844   \\
    \bottomrule
  \end{tabular}}
  \vspace{-3mm}
\end{table}

\subsection{Orderliness Test}

\begin{figure}[!htp]
  \centering
  \vspace{-1mm}  
  \includegraphics[width=0.45\textwidth]{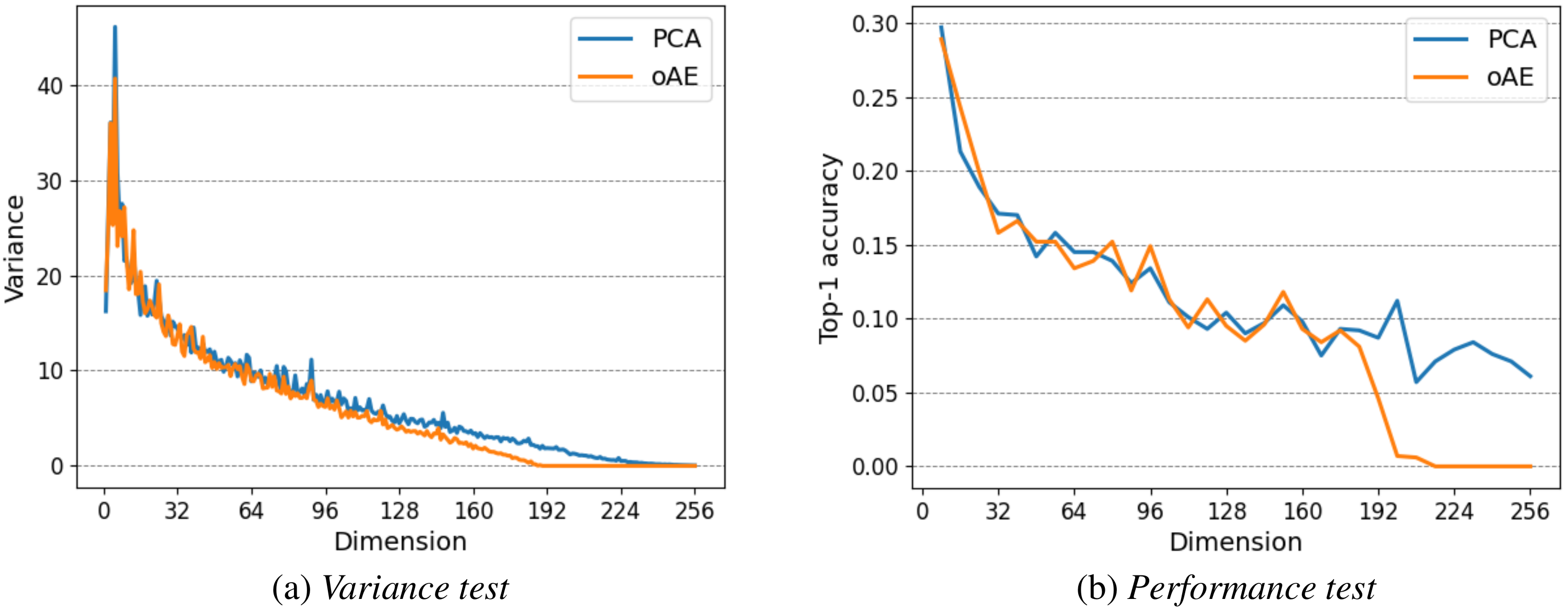}
  \vspace{-2mm}
  \caption{Orderliness test with PCA and oAE, using x-vectors in VoxCeleb1.}
  \vspace{-1mm}  
  \label{fig:order}
\end{figure}

In the second experiment, we test the orderliness of the features produced by oAE and PCA.
Figure~\ref{fig:order} illustrates (a) the variance of each dimension and (b) the
Top-1 accuracy with partial dimensions. In the partial-dimension test, each test involves
8 consecutive dimensions, resulting in 32 data points on each curve in Figure~\ref{fig:order}(b).

From Figure~\ref{fig:order}(a), one can observe that the variance of each dimension is quite similar with oAE and PCA,
and the leading dimensions show larger variations.
This demonstrates that oAE learns ordered representations as PCA. However, oAE is not a full copy of PCA. 
In particular, oAE performs even better than PCA on tail dimensions. The difference should
be attributed to the fact that oAE does not enforce parameter sharing between the encoder and decoder while PCA does.

From Figure~\ref{fig:order}(b), we can see that the partial-dimension performances of PCA and oAE are also similar and the trends are the
same: the leading dimensions are more important than the tail dimensions in the SID task.
This double confirms that the oAE model can learn ordered representations. Interestingly, this learning is based on
reconstruction error but works well on the SID task, suggesting some potential link between the reconstruction error and 
the cosine similarity used in SID.

\subsection{Binary Test}

In the third experiment, we test the OB codes derived from the obAE model.
In this experiment, the dimensionality of the latent representation is 256,
therefore the derived binary code involves 256 bits in total.
We will select the first $i \leq$ 256 bits to perform the test.

For a better comparison, we choose two alternative binary codes as references:
one is produced from the original x-vector with local sensitive Hashing (LSH)~\cite{charikar2002similarity,li2016binary}, denoted by LSH.
And the other one is produced from the PCA-transformed x-vector with LSH, denoted by PCA-LSH.
Results on the VoxCeleb1 and CN-Celeb.E datasets are reported in Table~\ref{tab:vox-binary} and Table~\ref{tab:cnc-binary}, respectively.

\begin{table}[!htp]
  \caption{Top-k Acc with three binary codes on VoxCeleb1.}
  \vspace{-2mm}
  \label{tab:vox-binary}
  \centering
  \scalebox{0.7}{
  \begin{tabular}{l|c|c|c|c|c|c|c}
    \toprule
    \multicolumn{2}{l|}{\textbf{Bits}}   & 20     & 40     & 80     & 120    & 160    & 256        \\
    \midrule
    \multirow{3}*{LSH}        & Top-1   & 0.094  & 0.273  & 0.543  & 0.684  & 0.759  & 0.847       \\
                              & Top-3   & 0.176  & 0.412  & 0.689  & 0.808  & 0.865  & 0.924       \\
                              & Top-5   & 0.227  & 0.481  & 0.747  & 0.851  & 0.898  & 0.945       \\
    \midrule
    \multirow{3}*{PCA-LSH}    & Top-1   & 0.126  & 0.350  & 0.599  & 0.709  & 0.768  & 0.847       \\
                              & Top-3   & 0.233  & 0.514  & 0.746  & 0.830  & 0.872  & 0.924       \\
                              & Top-5   & 0.297  & 0.588  & 0.799  & 0.870  & 0.904  & 0.945       \\
    \midrule
    \multirow{3}*{\textbf{obAE}} & Top-1 & 0.182 & 0.440 & 0.681  & 0.779  & 0.813  & 0.824       \\
                              & Top-3   & 0.316  & 0.616  & 0.822  & 0.890  & 0.911  & 0.913       \\
                              & Top-5   & 0.392  & 0.690  & 0.870  & 0.923  & 0.939  & 0.939       \\
    \bottomrule
  \end{tabular}}
\end{table}

\begin{table}[!htp]
  \caption{Top-k Acc with three binary codes on CN-Celeb.E.}
  \vspace{-2mm}
  \label{tab:cnc-binary}
  \centering
  \scalebox{0.7}{
    \begin{tabular}{l|c|c|c|c|c|c|c}
      \toprule
      \multicolumn{2}{l|}{\textbf{Bits}}      & 20    & 40    & 80    & 120   & 160   & 256      \\
        \midrule
        \multirow{3}*{LSH}        & Top-1     & 0.157 & 0.293 & 0.432 & 0.502 & 0.543 & 0.595    \\
                                  & Top-3     & 0.266 & 0.410 & 0.546 & 0.612 & 0.653 & 0.703    \\
                                  & Top-5     & 0.326 & 0.471 & 0.602 & 0.664 & 0.702 & 0.750    \\
        \midrule
        \multirow{3}*{PCA-LSH}    & Top-1     & 0.183 & 0.329 & 0.462 & 0.519 & 0.544 & 0.595     \\
                                  & Top-3     & 0.323 & 0.478 & 0.588 & 0.634 & 0.655 & 0.701     \\
                                  & Top-5     & 0.403 & 0.550 & 0.647 & 0.687 & 0.706 & 0.748     \\
        \midrule
        \multirow{3}*{\textbf{obAE}} & Top-1 & 0.197 & 0.353 & 0.495 & 0.551 & 0.579 & 0.588      \\
                                  & Top-3     & 0.357 & 0.518 & 0.639 & 0.688 & 0.708 & 0.719      \\
                                  & Top-5     & 0.446 & 0.594 & 0.705 & 0.743 & 0.762 & 0.774      \\
    \bottomrule
  \end{tabular}}
  \vspace{-1mm}  
\end{table}

It can be seen that PCA-LSH and obAE significantly outperform LSH when the code size is small.
However, as the code size becomes larger, this advantage will be gradually reduced and finally, 
the performance of the three codes tends to be the same.
This is not surprising as LSH can perfectly represent the dense vector if sufficient Hash functions are
used, according to the random matrix theory~\cite{charikar2002similarity}.
This means that the advantage any binary code may achieve is
in the regime of limited code capacity. And just in this regime, obAE shows a clear advantage over LSH, no matter whether PCA is employed.

\subsection{Bit Test}

\begin{figure}[!htp]
  \centering
  \includegraphics[width=0.45\textwidth]{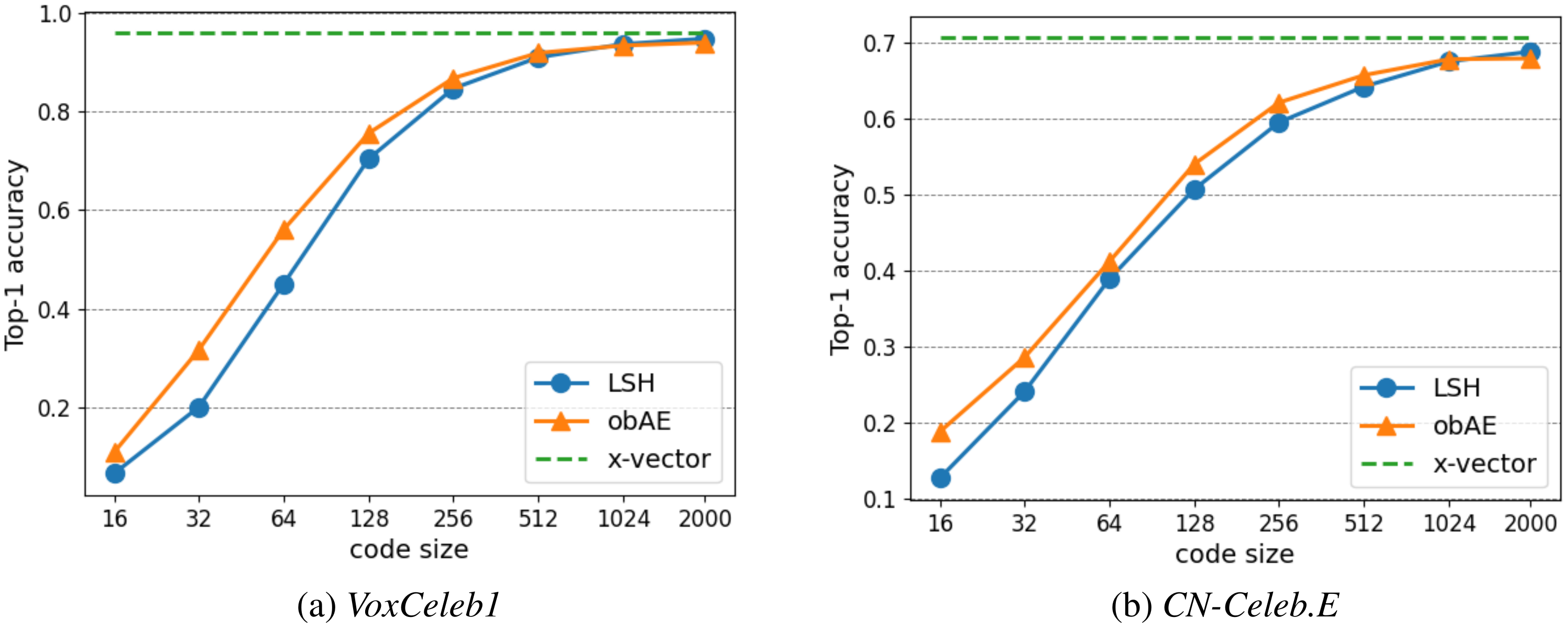}
  \vspace{-2mm}
  \caption{Bit test with LSH and obAE codes.}
  \label{fig:bits}
  \vspace{-2mm}  
\end{figure}

In the fourth experiment, we make a `bit test' to find out how many bits of the LSH and obAE codes can obtain comparable performance to the original x-vector.
To achieve this goal, we trained an obAE model with the dimensionality of the latent space set to 2,000, resulting in binary codes of 2,000 bits.
Once trained, the first $i \leq$ 2,000 bits are selected to make the performance test, as shown in Figure~\ref{fig:bits}.

It can be seen that as the code size increases, the Top-1 accuracy of the obAE codes gradually improves and eventually converges.
Compared to the original x-vector (8,192 bit) in Table~\ref{tab:base},
it seems that the OB codes could obtain a competitive performance when the code size reaches 1,000.
This indicates that there are lots of redundancy in the original x-vector, and binary codes may represent the full
information with less storage.

Finally, we see the performance of the obAE codes is consistently superior to the LSH codes, but ultimately they converge together.
Again, this indicates that LSH works well with sufficient code size, but obAE can achieve better performance with limited
code capacity. This feature is clearly attributed to the orderliness of the OB codes.

\subsection{Speed Test}

\begin{figure}[!htp]
  \centering
  \includegraphics[width=0.45\textwidth]{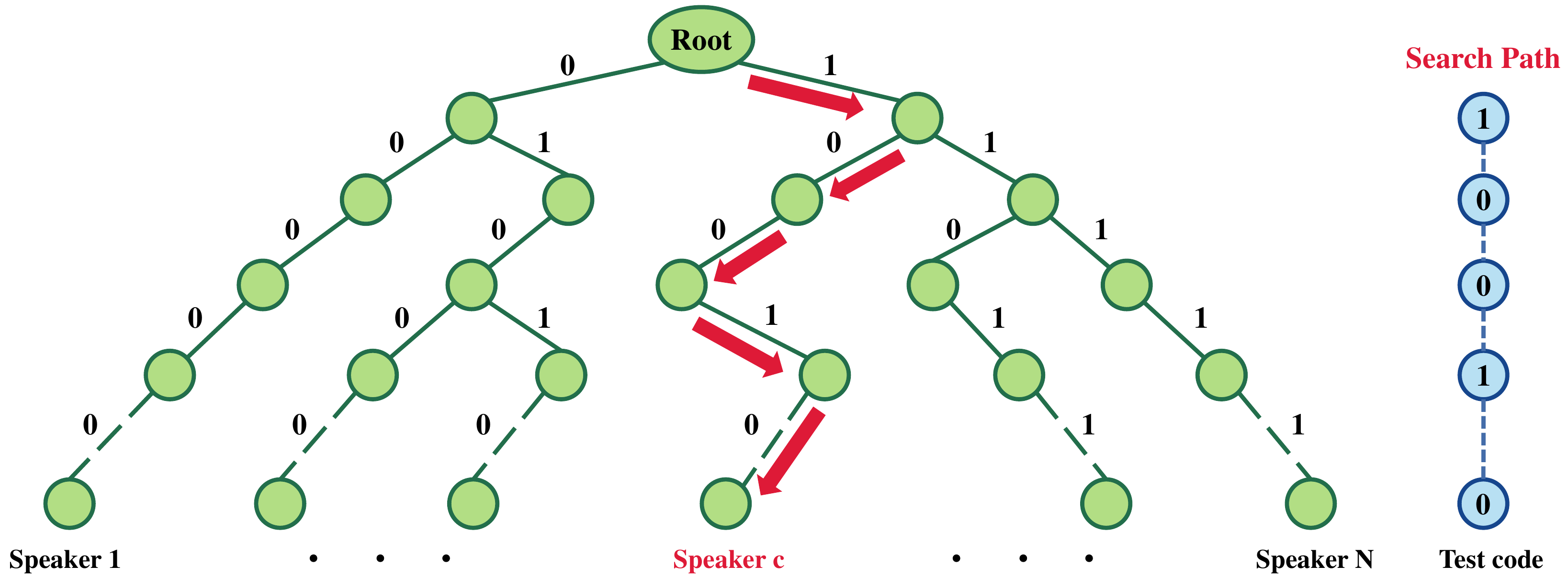}
  \vspace{-2mm}
  \caption{An illustration of the binary search tree.
  The leaf nodes correspond to enrollment speakers, and the red arrows show the search path of an identification task.}
  \label{fig:tree}
\end{figure}

As mentioned in Section~\ref{sec:intro}, the ordered and binary properties of the OB codes can be applied to gain fast retrieval.
Specifically, the OB codes form a binary tree, where the hierarchy is just the dimension index, as shown in Figure~\ref{fig:tree}.
If we assume that each leaf node of the tree corresponds to a single speaker,
then for any query OB code, a simple top-down traversal through the tree will reach the matched speaker, leading to very fast retrieval.
Most importantly, the complexity of this traversal-based retrieval is independent of the size of the speakers.
This means it is particularly suitable for search in a large population.

We further make an empirical test. To make the test simple, we
assume that each leaf node corresponds to a particular enrolled speaker and that each test OB code matches a leaf node.
This means that if a leaf node is reached by a test code, and the leaf node corresponds to the correct speaker of the query code, then a Top-1 hit is triggered.

We compare three retrieval modes: (1) linear search based on cosine distance with dense vectors produced by oAE; (2) linear search
based on Hamming distance with OB codes produced by obAE; (3) tree search with OB codes produced by obAE.
The results on VoxCeleb1 are reported in Table~\ref{tab:speed}.
It can be observed that with almost the same Top-1 accuracy, tree search obtains a remarkable advantage in speed:
it is about 1,300 times faster than the linear search based on cosine distance and 450 times faster than the linear search based on Hamming distance.
This demonstrates that OB codes possess great potential in large-scale retrieval tasks.

\begin{table}[!htp]
  \centering
  \caption{Speed test with linear cosine search, linear Hamming search, and binary tree search. `Speed' means the search time in millisecond and `Top-1' represents Top-1 accuracy.}
  \vspace{-2mm}
  \label{tab:speed}
  \scalebox{0.7}{
    \begin{tabular}{l|l|cc|cc|cc}
      \toprule
     \multirow{2}{*}{Code}& \multirow{2}{*}{Distance}& \multicolumn{2}{c|}{32 dims/bits}  & \multicolumn{2}{c|}{40 dims/bits}  & \multicolumn{2}{c}{48 dims/bits}    \\
                                     &     &  Speed     &  Top-1      &  Speed    &  Top-1       &  Speed     &  Top-1       \\
      \midrule
      Dense   & Cosine                    & 52.89      & 0.950       &  53.17    & 1.000        &  51.87     & 1.000        \\
      \midrule
      OB  & Hamming                   & 18.97      & 0.950       &  19.84    & 0.981        &  19.98     & 1.000        \\
      \midrule
      OB  & Binary tree               & 0.04       & 0.950       &  0.05     & 0.981        &  0.07      & 1.000        \\
      \bottomrule
    \end{tabular}}
\end{table}

\section{Conclusions}

In this paper, we presented an ordered binary AE (obAE) model to produce ordered binary (OB) speaker codes.
The model employs nested dropout to learn ordered representations and Bernoulli sampling to
get binary codes.
Extensive experiments were conducted to test the performance, orderliness, convergency, and speed of the ordered binary codes,
and the results showed that OB codes can bring remarkable speeding up, demonstrating the great potential of the new
approach in large-scale speaker identification tasks.
The future work will apply obAE to other speech processing tasks, e.g., acoustic model compression for speech recognition,
and investigate task-oriented supervision.



\bibliographystyle{IEEEtran}
\bibliography{mybib}

\end{document}